\documentclass[11pt,a4paper]{article}
\usepackage{graphics}
\usepackage{epsfig}
\usepackage{multicol}
\begin{document}
\vspace*{.1in}
\begin{center}
\begin{large}
{\textbf{DISCO: AN OBJECT-ORIENTED SYSTEM FOR}} \\[1ex]
{\textbf{MUSIC COMPOSITION AND SOUND DESIGN}}

\vspace{2ex}
Hans G.~Kaper,$^1$
Sever Tipei,$^2$ and
Jeff M.~Wright$^3$
\end{large}

\begin{footnotesize}
$^1$ MCS Division,
Argonne National Laboratory,
Argonne, IL 60439, USA
(\texttt{kaper@mcs.anl.gov}) \\
$^2$ School of Music,
University of Illinois at Urbana-Champaign,
Urbana, IL 61801, USA
(\texttt{s-tipei@uiuc.edu}) \\
$^3$ 
University of Illinois at Urbana-Champaign,
Urbana, IL 61801, USA
(\texttt{jmwrgh1@uiuc.edu})
\end{footnotesize}
\end{center}

\vspace*{1ex}
\begin{quote}
{\textbf{Abstract.}}
This paper describes an object-oriented approach to music composition
and sound design.  The approach unifies the processes of music making
and instrument building by using similar logic, objects, and procedures.
The composition modules use an abstract representation of musical data,
which can be easily mapped onto different synthesis languages or a
traditionally notated score.  An abstract base class is used to
derive classes on different time scales.  Objects can be related to act
across time scales, as well as across an entire piece, and relationships
between similar objects can replicate traditional music operations
or introduce new ones.  The DISCO (Digital Instrument for Sonification
and Composition) system is an open-ended work in progress.
\end{quote}

\vspace*{1ex}
\begin{multicols}{2}

\subsubsection*{1.~~INTRODUCTION}
The compositional process is based on the assumption that aural 
events can be ordered in time: a musical composition represents a 
trajectory in sound space.  The composer controls the structure, 
if not the details, of the trajectory and thus the nature of the
composition.  The control takes the form of an algorithm---a set 
of rules governing the nature of the objects, their evolution, and 
their interrelations---which defines the musical composition.  
Composing thus means defining objects and relating those attributes
that yield a desired trajectory in sound space.

The object-oriented paradigm and the software implementation we 
describe here reflect this point of view.  They also provide a way 
of merging two activities which, traditionally, have been considered
separate: writing music and building instruments.  With the exception
of Harry Partch~[Partch, 1960], who built actual instruments responsive
to his music's structure (based on ratios), and Xenakis~[Xenakis, 1993],
who used stochastic distributions to generate the structure
of computer-generated sounds as well as large scale textures, few
composers have shown an interest in combining these two areas.  
The system presented here treats both activities in a uniform way
by using similar logic, objects, and procedures.  The software
modules for music composition and sound design are consistently and
comprehensively interconnected.  The resulting code, currently
referred to as DISCO (Digital Instrument for Sonification and
Composition), is a work in progress.  The system was used recently
by one of the authors for the composition of a piece for violin
and computer-generated tape~[Tipei, 2000].

\subsubsection*{2.~~OBJECTS AND PROPERTIES}
The composition modules use an abstract representation of musical data, 
which can easily be mapped onto different synthesis languages or, as 
the case may be, a traditionally notated score.  This is achieved by 
defining ``Instrument'' and ``Property'' classes in response to the 
requirements of the target output.

The Instrument class is essentially a collection of properties
that define all of an instrument's control parameters.  A very
simple instrument might be defined by the properties ``Start Time,''
``Duration'' and ``Pitch.'' Each property is stored in a table,
which is indexed by a string identifier.  The Instrument class
includes the methods describing the manner in which the instrument's
output is to be generated.  Note that the Instrument class does not
necessarily correspond to any actual instrument, but serves rather
as an abstraction for defining the properties of a given sound object.

The Property class enables us to easily classify the different
properties of a sound object.  A composition would likely contain a
number of sound objects sharing certain properties, such as
``Start Time.''  In this case, the advantages of the polymorphic nature
of the system become evident, as one can work with these properties
without knowing the type of instrument.  The Property class
also incorporates methods to check for the correct type of input data.
For example, many instruments share the property ``Pitch,'' which may be
represented as a floating-point frequency value, as an integer that
indexes a tuning table, or as a string spelling the name of a note.

\subsubsection*{3.~~TIME SCALES}
The perception of aural events and their organization in larger
structures points to the existence of time scales associated with
particular objects.  We mention, in order of increasing magnitude,
the time scales of audio frequencies and of frequency and amplitude 
modulations, which affect partials and sounds; the time scales associated 
with melodic phrases, chordal aggregates and more complex textures; the 
time scales of larger formal units, such as sections and movements; and 
the time scale associated with an entire piece~[Kaper, 1999a].

An abstract base class, ``Event,'' is used to derive classes on different 
time scales. The Event class has a relatively simple structure, which 
is defined by three attributes: start time, duration and name. 
Subclasses are derived from it in response to particular needs.

An event may contain other events and thus become a ``Compound Event.''
An entire piece is the most inclusive compound event.  At the other
extreme are the ``Atomic Events,'' which do not include other events.  
Partials in a sound or the graphic symbol of a note in a printed score 
are examples of atomic events.  ``Sections,'' ``Phrases,'' ``Motives,''
``Chords'' and ``Aggreggates'' are compound events which contain events
of shorter or equal duration and may be themselves part of larger
structures---of other compound events.


Besides the three inherited attributes (start time, duration and name),
the derived classes have the property that they can be related to other
similar classes or to classes of finer or coarser granularity.  The type
of a class, as well as its potential relations to other classes, are
reflected in the class's name.

Relationships or associations can act across time scales.  An 
example is the congregation of partials into sounds, of sounds into 
chords or melodic gestures, and of sections into a composition.  Also,  
more sophisticated relationships can be established between objects 
at different time scales and/or different locations in the piece.  
For example, the presence of a sound with a particular spectral envelope 
may trigger the assignment of a specific chord in a remote section 
of the piece.

Relationships between similar objects can replicate traditional music 
operations, such as transposition, inversion, and retrograde of a group of 
sounds, augmentation/diminution of durations or pitch intervals, chord
inversion or other rearrangements of sounds in a chord, etc.

\subsubsection*{4.~~HIGHER LEVEL OF ABSTRACTION}
``Generator'' classes provide the composer with the ability to generate
events based on some specific algorithm.  They are designed to serve
across time scales and can be of a generalized or specific type.  For example,
a simple random generator can create ``NumberProperty'' objects, which
can be assigned any property of an instrument or event that is derived
from the NumberProperty class.  A specific generator to create only
events of a certain type can be obtained by combining several simple
generators into an ``Event Generator.''  One such utility, already in
place, is designed to select the number of partials within a sound,
the number of sounds in a cluster, or the number of sections in a
piece according to a selected probability distribution.  Another
utility, the ``Envelope'' class, also in place, reads an envelope and
interpolates values as necessary, thus giving the user control over
the shape of events on various time scales.  Still other classes enable
the user to assign values manually from a list of possibilities or by 
using a script.

We intend to design a number of common algorithmic composition
techniques as Generator classes to implement customized algorithmic
techniques of the composer's design.
These classes will be extendible and can be used by themselves,
as well as in combination.

\subsubsection*{5.~~METHODS AND APPLICATIONS}
The type of classes and the methods to relate them are determined 
by the type of music the user wishes to compose.  Objects like 
``Melody,'' ``Chord'' and ``Rhythm,'' and methods such as ``Canon'' and 
``Chorale'' anticipate a traditional composition; ``Markov,''
``Stochastic'' and ``Heterophony'' show a different bend.
While the initial emphasis was on less-than-traditional modes
of composing, the system has acquired a much wider scope and
now supports traditional, as well as nontraditional thinking.
In addition, it supports sound design for scientific sonification---the
faithful rendition of complex data sets in sounds~[Kaper, 1999b].
The DISCO system is a truly open-ended work in progress, which
is continuously being enriched with new classes and methods.

Among the first utilities developed for the DISCO system
was the ``Matrix'' class.
It was designed to enable the choice of a start time
and a duration for each section in a Manifold Composition
according to certain probability distributions.
A Manifold Composition is essentially a collection of variants
of one and the same piece, differing in details but with a similar
overall structure~[Tipei, 1989].
The differences between the variants are the result of stochastic choices.
We briefly explain how the Matrix class was used to construct
the probability matrices for the choice of start times and durations.

Suppose there are $n+1$ time marks in the piece
(including the start time and end time).
The start time of each section is supposed
to coincide with one of the time marks. The end time of the
piece cannot be the start time of a section, so there are $n$
possible start times; we label them $t_1$ through $t_n$.
Each time mark $t_j$ has a certain weight $q_j$ associated with it,
which measures the likelihood of the time mark becoming the start
time of a section.
Suppose there are $m$ possible sections, labeled $s_1$ through
$s_m$.  Each section $s_i$ has a certain (relative) weight $w_i$
associated with it; furthermore, $s_i$ has a certain probability
$p_{ij}$ to become active at the time mark $t_j$.
Using the Matrix class, a probability matrix $P$ is constructed
with $m$ rows and $n$ columns.
The elements of $P$ are
$$
  P_{ij}
   =
   \frac{\sum_{k=1}^{i}\sum_{l=1}^{j} w_k p_{kl} q_l}
        {\sum_{k=1}^{m}\sum_{l=1}^{n} w_k p_{kl} q_l} , \quad
   \begin{array}{l}
   i = 1, \ldots\,,m , \\
   j = 1, \ldots\,,n .
   \end{array}
$$
Then $P_{ij}$ is the probability that section $s_i$ will start
at the time mark $t_j$.
Once the start times have been chosen, the duration $d_i$
of each section $s_i$ is determined from a probability matrix $Q$,
which is constructed in a similar manner.

The matrices $P$ and $Q$ are dynamically adjusted.
Once a start time and a duration have been assigned
to a particular section, adjustments are made to diminish
the probability that any other section is selected
during the same time interval or at nearby times.

The Matrix class enables the assignment of events in any order,
not necessarily as they appear in the piece---a reflection of
the way most human composers work.  The class has the potential of 
correlating various rationales leading to a particular selection, and its
methods can be used in connection with any parameter values and intervals
of any event.   Not only sections in a piece can be defined this way,
but also sounds in a section, chords and motives in a section, etc.
A logical step will be to combine the matrices for the selection
of start times and durations into one three-dimensional matrix and, 
eventually, to include all parameters in a single multidimensional matrix.
Any one choice will then be the result of a combination 
of all available criteria and will determine all aspects of an event.  
Finding the appropriate data representation for such a multidimensional 
matrix, however, is not trivial---especially in C++.

\subsubsection*{6.~~INTERFACES}
All the basic classes described here have been implemented in C++.
However, even for experienced programmers, C++ is a difficult language,
and although some composers are excellent programmers, we cannot assume
that all composers are willing to spend the time and effort
to become proficient in C++.  For this reason, most C++ classes have
an analagous interface in Python, an interpreted high-level
object-oriented language that is considerably easier
to learn than C++~[Lutz, 1996; Beazley, 1999].
The choice of language is left to the user.

The wrapper code that allows the C++ classes to be used as Python classes
is generated by SWIG~[Beazley, 1996], which automates the process of combining
C and C++ code with higher-level languages such as Python, Perl and Tcl.
Although Python is currently the only language supported by the system,
it is relatively simple to generate wrappers for Perl and Tcl.

\subsubsection*{7.~~CONCLUSION}
In this paper we have described an object-oriented system
for music composition and sound design.
The object-oriented approach has the advantage
that one can easily add different classes and/or 
methods taylored to a particular composition
or aesthetic.  The code is like an open-ended work in 
progress, which invites the creation of structures
and relationships between sounds not yet employed.

\subsubsection*{ACKNOWLEDGMENTS}
\begin{small}

This work was supported by the Mathematical, Information, and
Computational Sciences Division subprogram of the Office of
Advanced Scientific Computing Research, U.S. Department of
Energy, under Contract W-31-109-Eng-38.
\end{small}

\subsubsection*{REFERENCES}
\begin{small}
Partch, H. 1960.
\textit{Genesis of a Music; An Account of a Creative Work,
 Its Roots and Its Fulfillments}, New York, Da Capo Press,
Second Edition (1974).

\noindent
Xenakis, I. 1992.
\textit{Formalized Music, Thought and Mathematics in Music},
Revised Edition, Pendragon Press, pp.~289--293.

\noindent
Tipei, S. 2000.
``AntiPhan'' for Violin and Computer-Generated Tape
(unpublished).

\noindent
Kaper, H. G. and Tipei, S. 1999a.
``Formalizing the Concept of Sound,''
Proc.\ Int'l.\ Computer Music Conference,
Beijing, China, pp.~387--390.

\noindent
Kaper, H. G., Tipei, S., and Wiebel, E. 1999b.
``Data Sonification and Sound Visualization,''
\textit{Computing in Science and Engineering},
Vol.~1, No.~4, pp.~48--58.

\noindent
Tipei, S. 1989.
``Manifold Compositions: A (Super)Computer-Assisted Composition Experiment
in Progress,''
Proc.\ Int'l.\ Computer Music Conference,
Columbus, Ohio, pp.~324--327.

\noindent
Lutz, M. 1996.
\textit{Programming Python},
O'Reilly \& Associates.

\noindent
Beazley, D. 1999.
\textit{Python Essential Reference},
New Riders.

\noindent
Beazley, D. 1996.
``SWIG: An Easy to Use Tool for Integrating Scripting Languages with C and C++,''
Presented at the 4th Annual Tcl/Tk Workshop, Monterey, Cal.
(http://www.swig.org/papers/Tcl96/tcl96.html)
\end{small}

\end{multicols}
\end{document}